\documentstyle[aps,epsfig,multicol]{revtex}
\begin{document}
\draft
\author{B.~All\'es$^{\rm a}$, J. J. Alonso$^{\rm b}$, C. Criado$^{\rm
b}$, M. Pepe$^{\rm a}$}

\address{$^{\rm a}$Dipartimento~di Fisica, Universit\`a di Milano-Bicocca
      and INFN Sezione di Milano, Milano, Italy}

\address{$^{\rm b}$Departamento de F\'{\i}sica Aplicada~I, 
Facultad de Ciencias, 29071 M\'alaga, Spain}

\title{Percolation properties of the 2D Heisenberg 
model}

\maketitle

$ $

\centerline{\it Abstract}
\begin{abstract}
We analyze the percolation properties of certain clusters defined on
configurations of the 2--dimensional Heisenberg model.
We find that, given any direction $\vec{n}$, in $O(3)$ space, 
the spins almost perpendicular to $\vec{n}$ form a percolating
cluster. This result gives indications of how the model can avoid a
previously conjectured Kosterlitz--Thouless phase transition at
finite temperature $T$.
\end{abstract}

\begin{multicols}{2}
\narrowtext


The classical Heisenberg model in 2 dimensions (2D) 
describes the behaviour of a system
of classical spins with short range ferromagnetic 
interactions~\cite{brezin}. 
The spins are placed at the sites of a 2D lattice. 
The physics of the model, which has been studied both through analytical 
calculations and Monte Carlo simulations, is defined by
equations that display a continuous $O(3)$ symmetry and it is subject
to the Mermin and Wagner theorem~\cite{mermin},
i.e.: there are no equilibrium states with broken symmetry. 

Perturbation theory (PT) indicates that the 2D Heisenberg model has 
a critical point at zero temperature~\cite{polyakov}. 
Moreover, from the 
field--theoretical point of view, the spin field carries a particle of 
non--zero mass $m$. This mass has been calculated by applying a 
Bethe ansatz technique~\cite{hasenfratz} and by making a partial 
use of PT. However, PT is constructed by studying small oscillations
around the trivial configuration (all spins parallel), 
a state which obviously violates the Mermin and Wagner theorem. Therefore
a problem raises concerning the validity of 
the above--mentioned analytical results and any other which relies on PT.

The model has also been studied by numerical simulations. 
Among other quantities (see for example~\cite{balog}), 
the mass $m$ and the magnetic susceptibility $\chi$ have been recently
measured with good precision. The mass is evaluated from 
the exponential decay of the 2--point correlation function $G(x-x')$.
The magnetic susceptibility is extracted from the sum 
$\chi\equiv \sum_x G(x-x')$. Some of the most recent numerical calculations
are the following.
In~\cite{kim,sokal} $m$ has been determined by
extrapolating the result from small lattice sizes and large temperatures to
smaller temperatures by using finite--size scaling. In~\cite{tomeu1} it
has been extracted from the 2--point function by using improved
actions and very high statistics. In both cases agreement with the
analytical calculation of $m$~\cite{hasenfratz} is found within 2--4\%.

Another scenario for the 2D Heisenberg model has been 
put forward in~\cite{stanley,seiler1,seiler2}. The model would undergo a 
Kosterlitz--Thouless (KT) phase transition at a finite temperature 
$T_{KT}$ and no massive particle would be carried by the spin field
in the low temperature phase.
This scenario is a mimic of what is known to happen in the
2D $O(2)$ or XY model~\cite{kosterlitz}.
If $T_{KT}$ is low enough, a numerical simulation
is not able to detect it (in~\cite{tomeu1,tomeu2} it was argued
that $T_{KT}$ is indeed much smaller than the typical temperatures 
utilized for thermalization in present--day simulations). 
In~\cite{schultka} a finite--size 
analysis of the helicity modulus is used to rule out
such a KT transition for $T > 0.1$. In~\cite{tomeu1,tomeu2}
it was also shown that the data for the correlation length and the
magnetic susceptibility for temperatures $T > 0.53 $ do not scale 
as the KT scenario predicts. 

In the present paper, we want to tackle directly the arguments 
of~\cite{seiler1,seiler2} where the percolation properties of certain
clusters are analyzed and, after assuming a set of hypotheses, it is
concluded that the magnetic susceptibility
diverges which is sufficient to prove that the mass $m$ vanishes. 

We realize the classical spins of the Heisenberg model by 3--component
scalar fields of modulus 1 placed at each site $x$ of a square lattice, 
$\vec{\phi}_x$ with $\left(\vec{\phi}_x\right)^2=1$. 
These fields interact through
a nearest neighbour (n.n.) coupling and the hamiltonian can be written
as ($\langle x\, y \rangle$ stands for two generic n.n. sites)
\begin{equation}
 H \equiv \sum_{\langle x\, y\rangle} \vec{\phi}_x \cdot 
 \vec{\phi}_y \;.
\label{eq:h}
\end{equation}
The partition function at a temperature $T$ 
is $Z=\sum_{\{\vec{\phi}_x\}} 
\exp\left(H/T\right)$. The hamiltonian (\ref{eq:h}) 
and the partition function are invariant under $O(3)$ rotations.

In the following we 
recall the arguments of~\cite{seiler1,seiler2}.
Let us consider a configuration for this model thermalized at a given
temperature $T$. Let $\vec{n}$ be an arbitrary unit vector in
the internal space of the $O(3)$ symmetry. To any such a vector we
can associate various types of clusters on the configuration. 
If $\cal{A}$ is one such cluster then its size, defined as the
number of sites contained in it, shall be denoted by $C_{\cal{A}}$. On the
other hand, its perimeter, defined as the number of sites along the
border of the cluster, will be called $B_{\cal{A}}$. If a set of
clusters completely cover the whole lattice volume with no overlap
then we say that we have a ``cluster system''.

The Fortuin--Kasteleyn clusters 
(hereafter called $\cal{F}$)~\cite{fortuin} are made of
sites connected by the bonds which survive the deletion process
performed with the probability
\begin{equation}
 P_{xy} \equiv \exp\left(\min\{0\,,\;-{2 \over T}
        \left( \vec{\phi}_x \cdot \vec{n} \right)
        \left( \vec{\phi}_y \cdot \vec{n} \right) \}\right) \;.
\label{eq:fk}
\end{equation}
The average size of the $\cal{F}$ clusters satisfies 
$\langle C_{\cal{F}} \rangle= \kappa\,\chi$ 
where $\chi$ is the magnetic susceptibility and $\kappa$
is a constant $\kappa < 1$. The brackets $\langle \cdot \rangle$
indicate average over configurations or equivalently the average
calculated with the Boltzmann weight of the partition function. 
In physical terms, $\cal{F}$ clusters are regions of correlated spins.
The set of $\cal{F}$ clusters form a cluster system.

Other clusters associated to an arbitrary unit vector $\vec{n}$ 
are the $\cal{H}^+$, $\cal{H}^-$ and $\cal{S}$
clusters. All n.n. spins on a thermalized configuration at a
temperature $T$ tipically satisfy
$\| \vec\phi_x - \vec\phi_y\| \leq \varepsilon$ with a parameter
$\varepsilon$ of order $O\left(\sqrt{2\,T}\right)$.
Then for any site $x$ the
scalar product $\left(\vec{\phi}_x \cdot \vec{n}\right)$ can be either
{\it a)} $\left(\vec{\phi}_x \cdot \vec{n}\right) > \varepsilon/2$, 
and $x$ belongs to an $\cal{H}^+$ cluster, or {\it b)} 
$\left(\vec{\phi}_x \cdot \vec{n}\right) < -\varepsilon/2$,
so $x$ belongs to an $\cal{H}^-$ cluster, or {\it c)}
$|\vec{\phi}_x \cdot \vec{n}| \leq \varepsilon/2$ 
and consequently $x$ belongs to 
an $\cal{S}$ cluster. In simple words, 
the $\cal{S}$ clusters are constituted by sites whose
spins almost lie in the plane perpendicular to $\vec{n}$;
on the other hand the $\cal{H}^+$ ($\cal{H}^-$) clusters 
contain sites whose spins are almost parallel (antiparallel) 
to $\vec{n}$. The set of $\cal{H}^\pm$ and
$\cal{S}$ clusters form a cluster system. 

By using a variant of the $O(3)$ model that includes a Lipschitz 
continuity condition, $\|\vec\phi_x-\vec\phi_y\|\leq\delta$ for all
n.n. $x$, $y$ (this condition does not change the physical
properties of the model as long as $\delta > \varepsilon$), 
it is possible to prove that 
the $\cal{H}^{\pm}$ clusters lie entirely
inside the $\cal{F}$ clusters~\cite{seiler1,seiler2}. Moreover no
$\cal{H}^{+}$ cluster has a common frontier with any
$\cal{H}^{-}$ cluster, they are always separated by spins belonging
to $\cal{S}$ type clusters.

Following Ref.~\cite{seiler1,seiler2}, 
three hypotheses are now necessary to prove that the magnetic
susceptibility diverges. The first one states the impossibility
to have simultaneous percolation of two or more disjoint
clusters in a cluster
system defined on the 2D Heisenberg model.
This fact is known to be true in several models~\cite{aizenman}.
The second hypothesis extends the validity of the
Mermin and Wagner theorem to hamiltonians which, 
like the one including the Lipschitz condition, 
are not differentiable in the fields. These
two conditions altogether prevent the $\cal{H}^\pm$ 
clusters from percolating.

The third hypothesis is to assume
that the $\cal{S}$ type clusters do not percolate either. 
Consequently none of the $\cal{H}^\pm$ or $\cal{S}$ clusters
percolate. When a cluster system does not contain a percolating 
cluster then at least two kinds of clusters have a divergent
average size~\cite{bunde}.
In the cluster system of $\cal{H}^\pm$ and $\cal{S}$, 
this result means that either 
$\langle C_{\cal{H}^+} \rangle = 
\infty$ or $\langle C_{\cal{H}^-} \rangle = \infty$ or both
(actually both, to avoid the breaking of the second hypothesis).
Moreover, as the
$\cal{H}^{\pm}$ clusters are entirely included inside 
the $\cal{F}$ clusters,
these clusters must also have a divergent average size. Therefore
the magnetic susceptibility $\chi$ diverges and the theory
is massless in clear contrast to the calculation of~\cite{hasenfratz}.

The above conclusion can be avoided if at least one 
of the three hypotheses fails.
In particular we have checked the third one finding that
there does exist a $\cal{S}$ cluster which percolates. 
In the present paper we report the results of a numerical
simulation of the Heisenberg model at $T=0.5$ on very large
lattices to give evidence that this is indeed the case at that 
temperature. We 
describe how this percolation takes place in the system, showing that
a percolating $\cal{S}$ cluster is found for every direction
$\vec{n}$ in the symmetry group space $O(3)$. 
Moreover all (percolating or not) $\cal{S}$ and $\cal{H}^\pm$ 
clusters present a high degree of roughness. More specifically,
every spin in a cluster has on average one n.n. lying on the
border of the cluster,
(on a 2D square lattice each spin has
four n.n.). In the following we describe the technical
details of our simulation and give the results obtained.


We have simulated the system described by the hamiltonian in 
Eq.(\ref{eq:h}) on several lattice sizes $L^2$ with 
$L=1024$, 1250, 1550 and 2048 at a temperature 
$T=0.5$. The measurements and cluster analysis have been performed on a
sample of between 1000 and 2000 configurations which were separated
by 200 overheatbath updating steps~\cite{petronzio}, (this updating
method applies a usual heatbath step to the dynamical variables --the
angle formed by the spin field $\vec{\phi}_x$ and the sum of their
n.n.-- and fixes the other variables by maximizing the distance
between the old and the new updated fields $\vec{\phi}_x$ in order to
hasten the decorrelation).

The percolation of clusters is a well--defined concept on infinite
lattices. Obvious computer limitations force us to work on finite
volume systems. We have overcome this difficulty by working with a
large ratio $L/\xi$ where $\xi$ is a typical correlation length of 
the system. In this way we expect that the would--be percolating
cluster will show up as a very large cluster, much larger than 
the others. We have chosen our simulation parameters in such a way
that $L/\xi\geq 4.5$, see Ref.~\cite{apostol}.

\begin{figure}[!]
\centerline{\epsfig{file=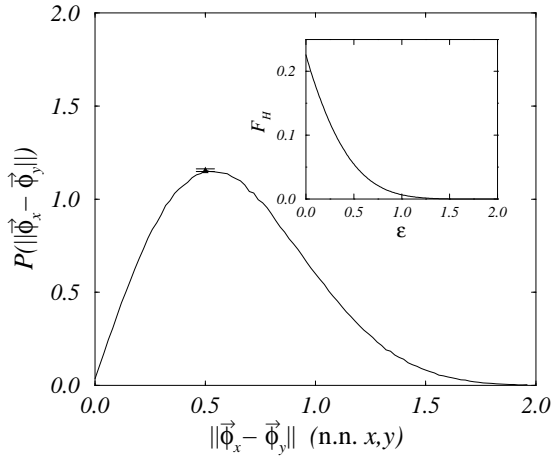,angle=-90,width=0.4\textwidth}}
\caption{Probability distribution of 
$\| \vec\phi_x - \vec\phi_y\|$ for all n.n. $x$, $y$ on a lattice
of size $L=1024$ at $T=0.5$. In the inset: fraction of bonds 
that connect $\cal{H}^+$ and $\cal{H}^-$ sites as a function of
$\varepsilon$.}
\label{fig1}
\end{figure}

The first task is to decide the value for $\varepsilon$ to be used in the
construction of clusters. $\varepsilon$ should not be too large in order
to avoid a certain (but not instructive for our problem) percolation
of the $\cal{S}$ clusters, the estimate
$\varepsilon\approx\sqrt{2\,T}$ being an appropriate value. 
In Fig.~\ref{fig1} we show the probability distribution of values 
of $\| \vec\phi_x - \vec\phi_y\|$ for all n.n. $x$, $y$ on a lattice
of size $L=1024$ at $T=0.5$. The error bars in this and in the
subsequent figures are very small: we show only the error bar of the
data point on the top of Fig.~\ref{fig1}.
Motivated by this distribution we have taken $\varepsilon=1.05$ 
which satisfies the estimate $\varepsilon\approx\sqrt{2\,T}=1$. We have checked
that, by using this value of $\varepsilon$, the $\cal{H}^+$ and
$\cal{H}^-$ clusters touch each other only in a $0.5$\% of all
bonds and this is shown in the inset of Fig.~\ref{fig1} where we give 
the fraction of bonds which
join $\cal{H}^+$ and $\cal{H}^-$ clusters, $F_H$,
as a function of $\varepsilon$. At $\varepsilon=1.05$ we will
already find evidence for the existence of 
percolating $\cal{S}$ clusters. For smaller epsilons one
cannot longer prove that the $\cal{H}^\pm$ clusters lie
entirely inside the $\cal{F}$ clusters, which is an essential ingredient for
the arguments given in~\cite{seiler1,seiler2}. We
could choose smaller values of $\varepsilon$ only at lower temperatures.

We have taken several choices for the vector $\vec{n}$. In all cases
we observed exactly the same results. In this paper we 
report the figures obtained with $\vec{n}=(1,0,0)$. 

\begin{figure}[!]
\centerline{\epsfig{file=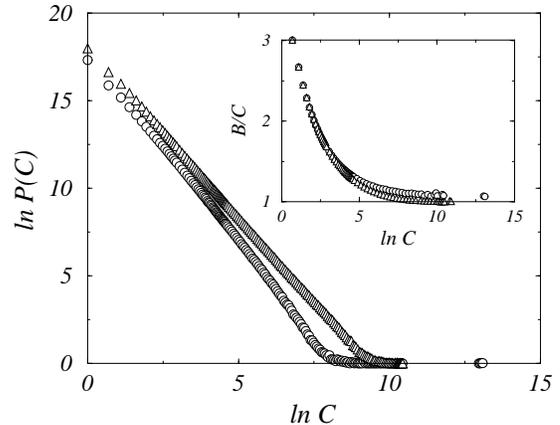,angle=-90,width=0.4\textwidth}}
\caption{Size distribution of $\cal{S}$ (circles)
and $\cal{H}^\pm$ (triangles) clusters on $L=1024$ at $T=0.5$ and
$\varepsilon=1$ on a log--log scale. In the inset: ratio
perimeter/size of clusters as a function of their size.}
\label{fig2}
\end{figure}

In Fig.~\ref{fig2} we show the distribution of 
sizes for the three types of clusters, $\cal{S}$ (circles)
and $\cal{H}^\pm$ (triangles). We present the data for each
cluster size $C$ up to sizes $C= 100$; for larger values of $C$
we show the results averaged over bins 
[$\ln C-\eta/2 , \ln C+\eta/2$] with $\eta\approx 0.5$. 
The quantity $P(C)\,{\rm d}C$ is proportional  
to the probability of finding a cluster with size between 
$C$ and $C+{\rm d}C$. The line with triangles 
is continuous and it ends at $\ln C=10.2$. On the other hand 
the curve with circles, which represents the $\cal{S}$ clusters, 
displays two parts: a continuous line 
ending at $\ln C=9.2$ and a separated point at $\ln
C\approx 13.1$. This isolated point is the one determined by the percolating
cluster. Although not explicitely depicted,
this point has an horizontal error bar which results in 
the slight horizontal spreading of circles. This error bar is a remnant of
the fact that on truly infinite lattices the size of the percolating
cluster must be infinite. Notice that the continuous part of the curve
with circles becomes steeper just before ending at $\ln C=9.2$: this
is another indication for the existence of a percolating
cluster~\cite{cambier} because it means that all clusters beyond 
some size prefer to be absorbed by the percolating $\cal{S}$ cluster.
This figure has been obtained by working on a lattice size 
$L=1024$ and choosing $\vec{n}=(1,0,0)$. Completely analogous results
are obtained with other sizes $L$ and any other unit vector $\vec{n}$.
We then conclude that there always exists a percolating
cluster of type $\cal{S}$ for any versor $\vec{n}$. 

\begin{figure}[!]
\centerline{\epsfig{file=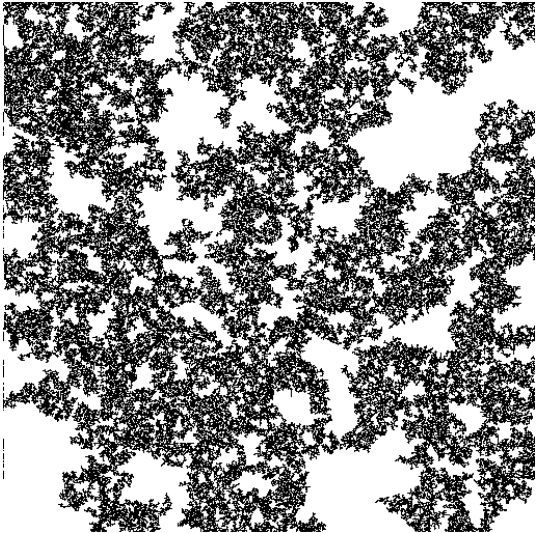,width=0.5\textwidth}}
\caption{Percolating $\cal{S}$ cluster (in black) from a 
thermalized configuration on a lattice of size $L=1024$.}
\label{fig3}
\end{figure}

In Fig.~\ref{fig3} an example of a type $\cal{S}$ percolating 
cluster taken from a single thermalized configuration is shown. 
The percolating sites are coloured in black.
To render the fine details of the clusters as clear as possible we
show only a piece of size $512\times 512$ taken out from a configuration 
on a lattice with $L=1024$. The figure corresponds to the
percolating cluster found in the plane perpendicular to
$\vec{n}=(1,0,0)$. It is clear from this figure that
the clusters present a high degree of roughness. 

The property of roughness is made explicit in the inset 
of Fig.~\ref{fig2}. The ratio ``perimeter of cluster''/``size of 
cluster'', $B/C$, is displayed against $\ln C$ for all $\cal{H}^\pm$
(triangles) and $\cal{S}$ (circles) clusters 
(irrespective of the fact that they
percolate or not). For bidimensional compact objects 
$B/C \propto 1/\sqrt{C}$ when $C \rightarrow \infty$. 
In our case this ratio tends to a constant (which is almost 1) 
when $C \rightarrow \infty$ and
this is an indicative that our clusters present a rough border, 
suggesting that they have a fractal structure. 

In the inset of Fig.~\ref{fig2} we see that the ratio $B/C$ goes
asymptotically to a constant which seems to be a bit 
larger for the $\cal{S}$ clusters than for $\cal{H}^\pm$. We think
that this is because the
$\cal{S}$ clusters surround the $\cal{H}^\pm$
clusters (see the introduction) and consequently they need to have an
additional boundary.

We have also studied the ratio among the size of the largest
(always percolating) $\cal{S}$ cluster, 
$M_{\cal{S}}\equiv\max\{C_{\cal{S}}\}$, and the size of the largest
$\cal{H}^\pm$, $M_{\cal{H}}\equiv\max\{C_{\cal{H}^\pm}\}$, 
as a function of the lattice size $L$. Let us call $R$
such ratio,
\begin{equation}
 R\equiv {\left\langle M_{\cal{S}} \right\rangle
         \over
          \left\langle M_{\cal{H}} \right\rangle}\; .
\label{r}
\end{equation}
If $R$ increases with $L$ for a fixed temperature, we
can reasonably infer that the above described percolation properties
survive the thermodynamic limit.
The data for $R$ and the maximum size of $\cal{S}$ and
$\cal{H}^\pm$ clusters are shown in Table~1 from the four 
lattice sizes we have used. They indicate that indeed the largest
$\cal{S}$ cluster keeps percolating when $L$ increases. Notice that
the percentage of lattice volume covered by the percolating cluster
does not vary appreciably with the lattice size $L$.

{{{\rm TABLE 1.} Ratio $R$ and maximum sizes of clusters 
as a function of the lattice size $L$.}
\vskip 2mm
{\centerline{
\begin{tabular}{|p{16mm}|p{16mm}|p{15mm}|p{15mm}|p{16mm}|} \hline
$L$ & 1024 & 1250 & 1550 & 2048 \\ \hline
$\left\langle M_{\cal{S}}  \right\rangle/L^2$ 
 & 0.437(10) & 0.429(4) & 0.433(7) & 0.424(4)  \\ \hline
$\left\langle M_{\cal{H}}  \right\rangle/L^2$
 & 0.019(3) & 0.015(2) & 0.011(1) & 0.0072(7)  \\ \hline
$R$ & 23(4) & 29(4) & 40(4) & 59(6) \\ \hline
\end{tabular}
}}}

\vskip 4mm


In conclusion, we have studied the percolation 
properties of several clusters which
can be defined on configurations of the 2D Heisenberg
model described by the hamiltonian in Eq.~(\ref{eq:h}). Of particular
interest are the clusters called $\cal{S}$. To define them one has to
introduce an arbitrary unit vector $\vec{n}$ in the symmetry group
space of the system, $O(3)$. When the spin $\vec{\phi}_x$ at the site
$x$ is almost perpendicular to $\vec{n}$, we say that this site
belongs to some $\cal{S}$ cluster. The term ``almost perpendicular''
depends on the temperature $T$ of the system (see above). For $T=0.5$
and working on rather large lattice sizes $L\geq 1024$, we have given
strong evidence  
that for every $\vec{n}$ one of these $\cal{S}$ clusters percolates on
each thermalized configuration of the system (see Fig.~\ref{fig2}).
These percolation properties seem to survive the thermodynamic limit 
(see Table~1). 

This is an important conclusion because if such clusters do not 
percolate then one can prove~\cite{seiler1,seiler2} that the system
carries no massive particle, contradicting 
the exact calculation of the mass gap performed for this
theory in Ref.~\cite{hasenfratz}. Our results exclude this
massless phase for $T > 0.5$.

We have also shown that, at least at $T=0.5$, all clusters 
present a high degree of roughness recalling a fractal structure,
(see inset in Fig.~\ref{fig2} and Fig.~\ref{fig3}).
It seems unlikely that, as suggested in~\cite{seiler1}, such a
dilute set of spins, almost lying on a plane, can enforce the system
to behave like an effective $O(2)$ model (see~\cite{koma}).

\vskip 6mm

We wish to thank Andrea Pelissetto for discussions and 
Julio Fern\'andez for a careful reading of the manuscript.
B.~A.~also thanks the warm hospitality at the Departamento de
F\'{\i}sica Aplicada~I of the M\'alaga University during the completion
of the paper.


\end{multicols}

\begin{references}
\bibitem{aizenman} See ~\cite{seiler2} and references therein.
Our formulation of this hypothesis is less general than the original one
appearing in~\cite{seiler2} but it is
still enough for the purpose of the proof.
\end{references}
\end{document}